\documentclass[aps,prb,twocolumn]{revtex4-1}
\usepackage{graphics}
\usepackage{subfigure}
\usepackage{epsfig}
\usepackage{epsf,epic}
\usepackage{color}
\usepackage{marvosym}
\usepackage{subfigure}
\usepackage{amsmath}
\usepackage{amssymb}
\usepackage{amsfonts}
\usepackage{wrapfig}
\usepackage{pstricks}
\usepackage{multirow}
\usepackage{color}
\usepackage{bm}
\usepackage{dcolumn}
\newcommand{\etal}{\textit{et al.\ }}

\begin{document}
\title{Native interstitial defects in ZnGeN$_2$}
\author{Dmitry Skachkov and Walter R. L. Lambrecht}
\affiliation{Department of Physics, Case Western Reserve University, 10900 Euclid Avenue, Cleveland, Ohio 44106-7079, USA}
\begin{abstract}
  A density functional study is presented of the interstitial Zn$_i$, Ge$_i$, and N$_i$ in ZnGeN$_2$. Corrections to the band gap are included by
  means of the LDA+U method. 
   The Zn and Ge interstitials are both found to strongly
  prefer the larger octahedral site compared to the two
  types of tetrahedral sites.
  The Zn  interstitial is found to be a shallow double donor
  but has higher energy than previously studied antisite defects. It
  has a resonance in the conduction band which is Zn-$s$ like.
  The Ge interstitial is an even higher energy of formation defect
  and also behaves as a shallow double donor but
  has also a deep level in the gap, corresponding  to a Ge-$s$ orbital character while
  the Ge-$p$  forms a resonance in the conduction band. 
  The nitrogen
  interstitial forms a split-interstitial configuration, as also
  occurs in GaN. Its electronic levels can be related to that of a N$_2$
  molecule. The defect levels in the gap correspond to the $\pi_g$-like lowest unoccupied molecular orbital (LUMO) of the molecule, which here becomes filled with 3 electrons in the  defect's neutral charge state. They are found to
  prefer a high-spin configuration in the $q=+1$ state. 
  The corresponding transition
  levels are obtained and show that this is an amphoteric trap level occurring
  in $+2$, $+1$, 0 and $-1$ charge states. The two possible sites for this
  split interstitial, on top of Zn or on top of Ge differ slightly in
  N$_2$ bond-length.   While the N$_i$ defects have the lowest formation energy among the interstitials, it is still higher than that of the antisites.
  Hence they are not expected to occur in sufficient concentration to
  affect the intrinsic Fermi level position. In particular, they do not
  contribute to the unintentional n-type background doping. 
\end{abstract}
\maketitle
\section{Introduction}
The II-IV-N$_2$ analogs of GaN, such as ZnGeN$_2$, ZnSnN$_2$,
Mg-IV-N$_2$ and Cd-IV-N$_2$ compounds have
recently received considerable attention for their potential to complement the group-III nitrides as opto-electronic materials.\cite{Lambrechtbook,Punya11,Quayle15,Paudel07,Peshek08,Paudel08,Paudel09,Punyaiwn,Atchara16,Sai17,Du,Zhu}
ZnSnN$_2$ has received attention as a potential
solar cell material composed entirely of earth-abundant and non-toxic
elements.\cite{Feldberg12,Lahourcade13,Quayle13,Fioretti15,Qin16,Deng15,Senabulya16}
ZnGenN$_2$ on the other hand
is very well lattice-matched to GaN and has almost the same band gap
but has a large type-II band-offset to it.\cite{Punya13} This provides interesting
opportunities for band gap engineering in heterojunctions, which
may improve efficiency of light-emitting diodes by wave-function
shaping.\cite{Han16} 

For all of the envisioned semiconductor opto-electronic type applications, doping and defect control are
crucial. Until now only a few papers have addressed these issues.
Chen \etal \cite{Chen14}  studied native defects in ZnSnN$_2$ and
Wang \etal \cite{Wang17} studied the possibility of p-type doping in ZnSnN$_2$
by means of alkali metals. 
Recently, we presented a first study of the native defects in
ZnGeN$_2$\cite{Skachkov16}  and the role of exchange defects.\cite{Skachkov16x}
However, in those  studies  we only paid attention 
to vacancies and antisites. To complement that paper, we here present
a study of native interstitial defects, Zn$_i$, Ge$_i$ and N$_i$. 
\section{Computational Method}
Density functional theory (DFT)\cite{Hohenberg-Kohn,Kohn-Sham} in the local density approximation (LDA)\cite{vonBarthHedin}
is used in this study  to calculate energy levels and total energies of the defects. The full-potential linearized muffin-tin orbital (FP-LMTO) method is used as band-structure method to solve the Kohn-Sham equations.\cite{Methfessel,Kotani10} 
The defect are modeled in a supercell with periodic boundary conditions.
A cell-size of 128 atoms or $2\times2\times2$ of the orthorhombic
primitive 16 atom cell is used.  
The structure of the defects is fully relaxed while keeping the
cell volume fixed. The Brillouin zone integrations are done using the
$\Gamma$-point only. 
The energy of formation of a defect
is calculated as
\begin{equation}
\begin{split}
E_{\rm for}(D,q)=E_{\rm tot}(D,q)-E_{\rm tot}(X)+ \\
+\sum_i\mu_i\Delta n_i +q(\epsilon_v+\epsilon_F+ V_{\rm align})+
E_{\rm corr} \label{eqefor}
\end{split}
\end{equation}
Here $E_{\rm for}(D,q)$ is the energy of formation of defect $D$ in charge 
state $q$,  $E_{\rm tot}(D,q)$ is the 
corresponding total energy of the supercell, 
from which we actually already subtracted the free atom energies, 
$E_{\rm tot}(X)$ is the supercell total energy of the perfect crystal 
calculated in the same size supercell. The chemical potentials of the elements involved in creating the defects
$\mu_i$ can vary in certain ranges depending on the growth conditions, as discussed in Ref. \onlinecite{Skachkov16}.
In the present case of interstitials $\Delta n_i$ is $-1$, because we add an atom. Thus the higher the chemical
potential, or, the richer the growth environment is in the atom we consider as interstitial, the lower its
energy of formation. The last term, represents the dependence on the chemical potential of the electron,
or Fermi level, which is considered here as an independently tunable parameter, although one can also
determine it self-consistently from the defect concentrations, which follow in turn from the
energies of formation, and the charge neutrality requirement.\cite{Skachkov16}
The alignment potential,\cite{Freysoldt09,Freysoldt14} $V_{\rm align}$ and
finite-size or image-charge correction terms,\cite{Komsa13,Kumagai14} $E_{\rm corr}$
are treated as in our previous work.\cite{Skachkov16} In particular,  we note
that in some cases we treat these using an effective charge
rather than the nominal charge with the effective charge determined by examining
the behavior of the electrostatic potential as function of distance from
the defect. 

Because the LDA is well-known to underestimate the band gaps 
(here $E_g^\mathrm{LDA}=1.9$ eV instead of experimental value of 3.4 eV) this can
sometimes lead to wrong interpretation of the defect levels, we used an LDA+U approach to correct the gaps
and the one-electron defect levels as described in our previous work.\cite{Skachkov16}
The LDA+U approach, while originally introduced for localized partially filled $d$- or $f$-shells,\cite{Anisimov91,Liechtenstein95} is applied here to Ge-$s,p$ and Zn-$s,p$ states and has the effect of
shifting the potential of these orbital energies by $\Delta V_i=U_i(\frac{1}{2}-n_i)$. Hence for empty states $n_i=0$
the shift is $U_i/2$. The chosen orbitals are dominant in the conduction band, in particular the conduction band minimum (CBM) at $\Gamma$
is mostly Zn-$s$ and Ge-$s$ like, while states at the Brillouin zone boundary in the lowest conduction band
have more Zn-$p$ and Ge-$p$ character. While these LMTO orbitals do not have zero occupation it is less than $\frac{1}{2}$
and hence an upward shift is obtained. The values used, $U_{\mathrm{Ge}-s}=U_{\mathrm{Zn}-s}=3.5$ Ryd, $U_{\mathrm{Ge}-p}=U_{\mathrm{Zn}-p}=2.4$ Ryd give a uniform shift of the conduction band with a gap of 3.4 eV in good agreement with experiment.
Within this approach, the one-electron defect levels will shift according to how much their wave function contain
these LMTO orbitals in their basis set expansion. Thus, defect levels 
will shift up in proportion to how much they are  ``conduction band derived'', while defect levels which are
mostly ``valence band derived'' will not shift. 
The energies of formation are corrected by including the shift of their corresponding occupied one-electron energies only. 
We do not fully include the LDA+U total energy expressions here because we are not dealing with partially filled
local orbital bands for which these expressions were derived.\cite{Liechtenstein95} Instead the LDA+U method here is used only as a tool to
create an upward shift of the potential on certain orbitals. This avoids the problem of also having to
calculate the reference chemical potential energies with a similar LDA+U Hamiltonian. 
We performed all the calculations at fixed lattice constants 
$a=6.376$, $b=5.518$, $c=5.263$ \AA, which were optimized for
the host material within the LDA.

\section{Results}
\begin{figure}
\includegraphics[width=8cm]{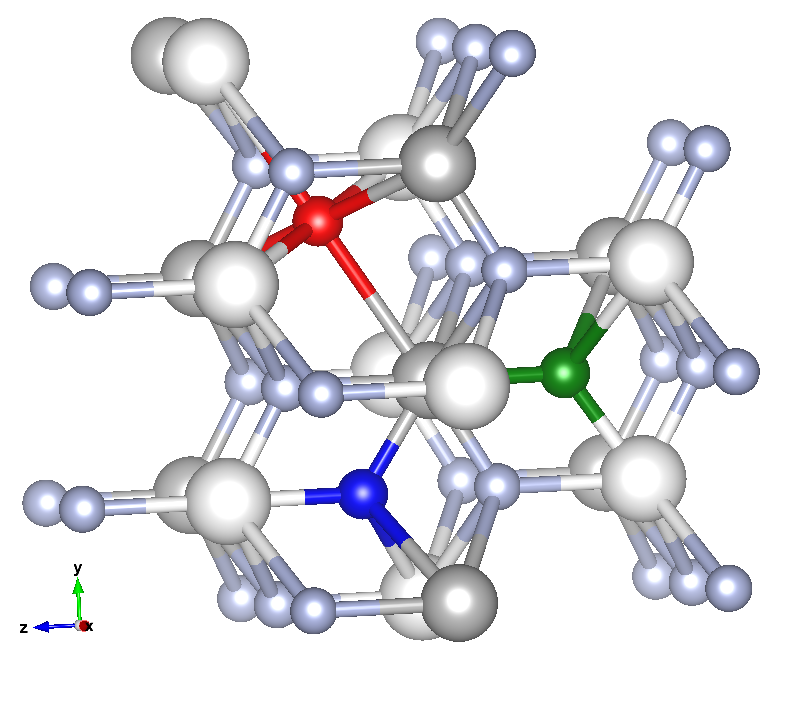}
\caption{(Color on-line) Three interstitials positions, octahedral (red) and
  two distinct tetrahedral (green and dark blue) in ZnGeN$_2$. Zn atoms represented by white spheres, Ge atoms by grey, and N atoms by light blue spheres.\label{i1_i2}}
\end{figure}

In the wurtzite structure, there are two high-symmetry interstitial sites,
the tetrahedral one, located vertically between a cation and anion
in the $c$-direction in two successive A layers of the ABAB stacking,
and an octahedral one, located in the big open hexagonal cross-section
channels, when viewed from the $c$-direction. In ZnGeN$_2$, there
are furthermore two types of tetrahedral interstitials depending on
whether it occurs between a Zn and N or a Ge and N (Fig. \ref{i1_i2}). We found in our relaxation calculations that both Zn and Ge strongly prefer the 
large octahedral sites. When we start out an interstitial in the tetrahedral
site it is unstable and moves toward the octahedral site. Our results
for both these cases thus correspond to the octahedral site.

\begin{figure}
\includegraphics[width=8.5cm]{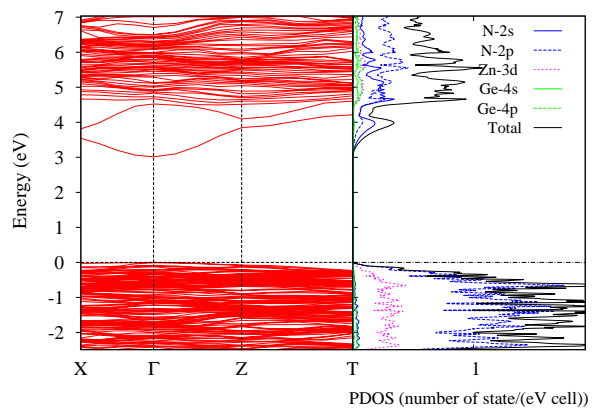}
\caption{(Color on-line) Band structure and projected density of states on nearest atoms 
to the ${\rm Zn}_{\rm i}$ in the $q=+2$ state obtained in the LDA+U approach.\label{bnds_Zn_i}}
\end{figure}

\begin{figure}[h]
    \includegraphics[width=8cm]{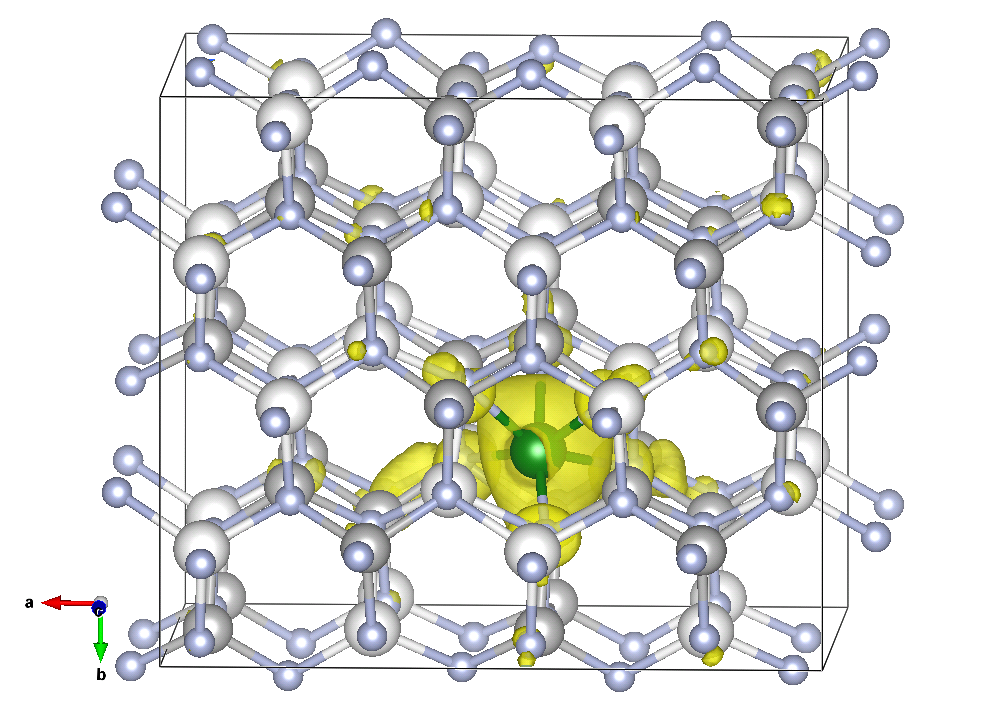}
    \caption{(Color online)  Wave function $|\psi|^2$
      for the resonance HOMO+2 in Zn$_{i}$ in the $q=+2$ charge state
      in LDA. Green sphere indicates Zn$_{i}$ atom. 
\label{wden_Zn_i_q_2}}
\end{figure}

We start by discussing the one-electron energy levels or
defect band structures of the supercells.
The band structure and partial density of states (PDOS) for Zn$_i$ are shown
in Fig. \ref{bnds_Zn_i} for the $q=+2$ charge state.

We can see that
there are no defect levels in the band gap, but a peak in density
of states slightly above the conduction band, and comparison to the
perfect crystal band structure indicates a resonance in the conduction
band. In the neutral and $q=+1$ charge states, we find the Fermi level  inside
the conduction band, but for the $q=+2$ state, it occurs at the
valence band maximum (VBM). This
indicates that Zn$_i$ is a shallow double donor defect. Fig. \ref{wden_Zn_i_q_2}
shows the wave function modulo squared for the second level above the conduction
band minimum in the the $q=+2$ state. This is a resonant level
and is calculated as the contribution to the charge density from a small
${\bf k}$-space region around $\Gamma$ and over a narrow energy range.
From its approximately spherical shape on the $s$-site, we conclude
it is mostly Zn-$s$ like.  The results shown here for the bands are
obtained in the LDA+U approach. Within the LDA, a similar result is obtained,
(shown in Supplementary Information) except that the gap is correspondingly smaller.
In other words the resonances in the conduction band shifted up along with the conduction band,
which is as expected from their Zn-$s$-like orbital  character.

\begin{figure*}[t]
\includegraphics[width=\linewidth]{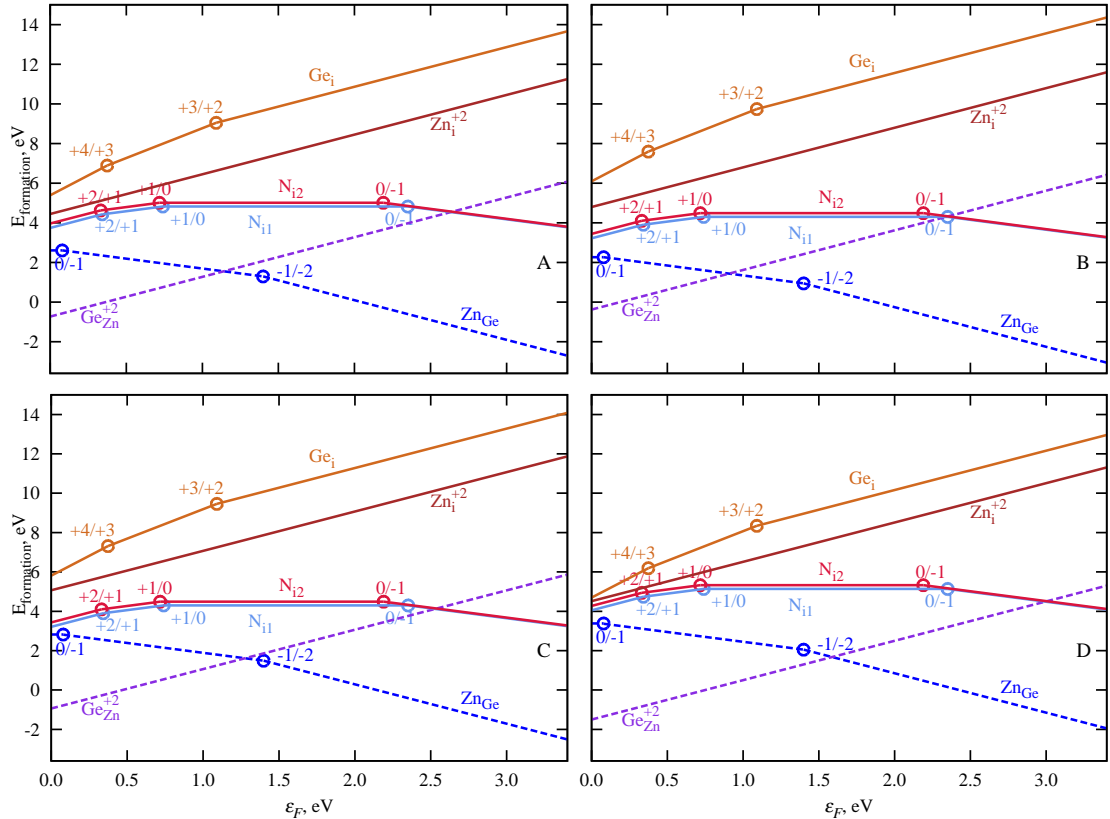}\vspace{-2cm}
 \caption{(Color online) 
   Formation energies for interstitials defects Zn$_i$, Ge$_i$, and N$_i$ along with the two most important (lowest energy) native defects Ge$_\mathrm{Zn}$, Zn$_\mathrm{Ge}$ (shown with dashed lines) for various chemical potential conditions, explained in the text.\label{E_for}}
\end{figure*}

\begin{table}
  \caption{Chemical potentials for ZnGeN$_2$ stability region (in eV).\label{tabchempot}}
  \begin{ruledtabular}
    \begin{tabular}{lcccc}
      & A & B & C & D \\ \hline
      $\Delta \mu_\mathrm{N}$ & -0.525 & 0 & 0 & -0.840 \\
      $\Delta \mu_\mathrm{Zn}$& 0      & -0.35 & -0.626 & -0.066 \\
      $\Delta \mu_\mathrm{Ge}$& -0.696 & -1.396 & -1.12 & 0 \\
    \end{tabular}
  \end{ruledtabular}
\end{table}

The corresponding energy of formation as function
of Fermi level is show along
with that of the other interstitials in Fig. \ref{E_for}.
The various panels here correspond to different chemical potential
conditions as defined in Fig. 1 of Ref. \onlinecite{Skachkov16},
summarized here in Table \ref{tabchempot}.
The lines AB and CD mark the boundary with the region where
Zn$_3$N$_2$ and Ge$_3$N$_4$ become the preferred phases, while the
line BC corresponds to N-rich condition, marking equilibrium with
N$_2$ molecules. One can see
that the Zn$_i$ has the lowest energy of formation for case A,
which is the most Zn-rich case, but in fact, does not vary very much because
the range of allowed Zn chemical potentials is rather restricted.
Similarly, Ge$_i$ has the lowest energy of formation for the most Ge-rich
condition D and N$_i$ has the lowest energy of formation both B and C. 
For comparison, this figure also includes the energies of formation of the
Zn$_\mathrm{Ge}$ and Ge$_\mathrm{Zn}$ antisites, which are the lowest
energy defects found in Ref. \onlinecite{Skachkov16}.
For the Zn$_i$, only the $q=+2$ charge state occurs in the gap.

\begin{figure}
\includegraphics[width=8.5cm]{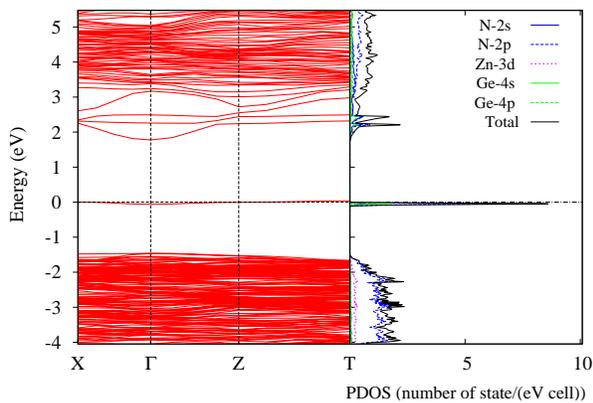}
\caption{(Color on-line) Band structure and projected density of states on nearest atoms 
to the ${\rm Ge}_{\rm i}$ in the $q=+2$ state obtained in LDA+U.\label{bnds_Ge_i}}
\end{figure}

\begin{figure}[h]
  \subfigure[]{
    \includegraphics[width=7cm]{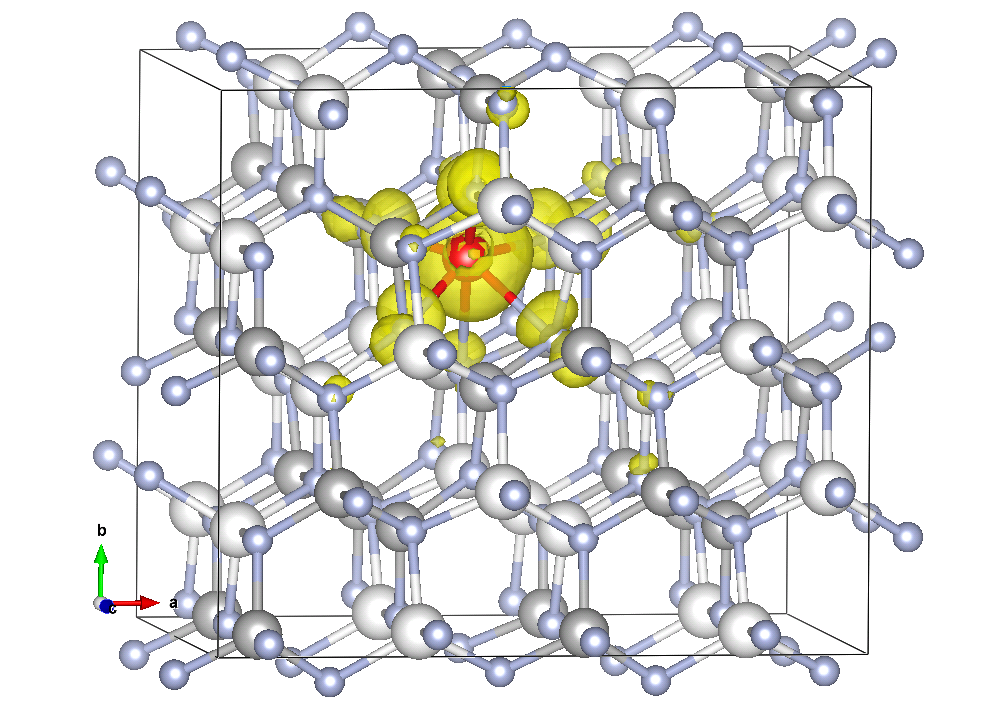}}
  \subfigure[]{
 \includegraphics[width=7cm]{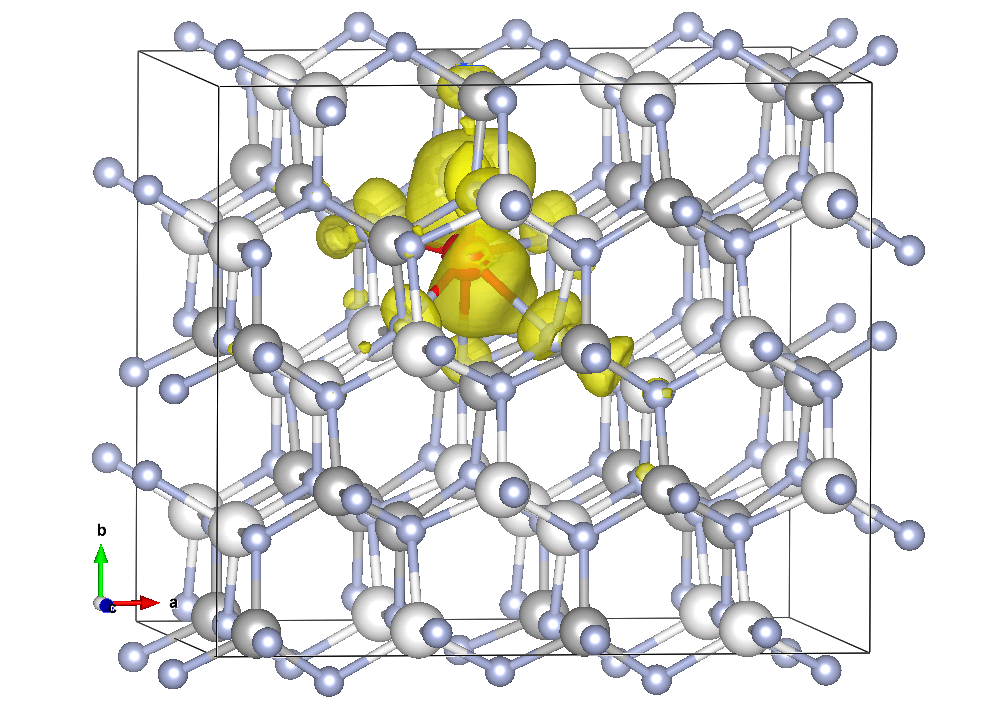}}
\subfigure[]{ \includegraphics[width=7cm]{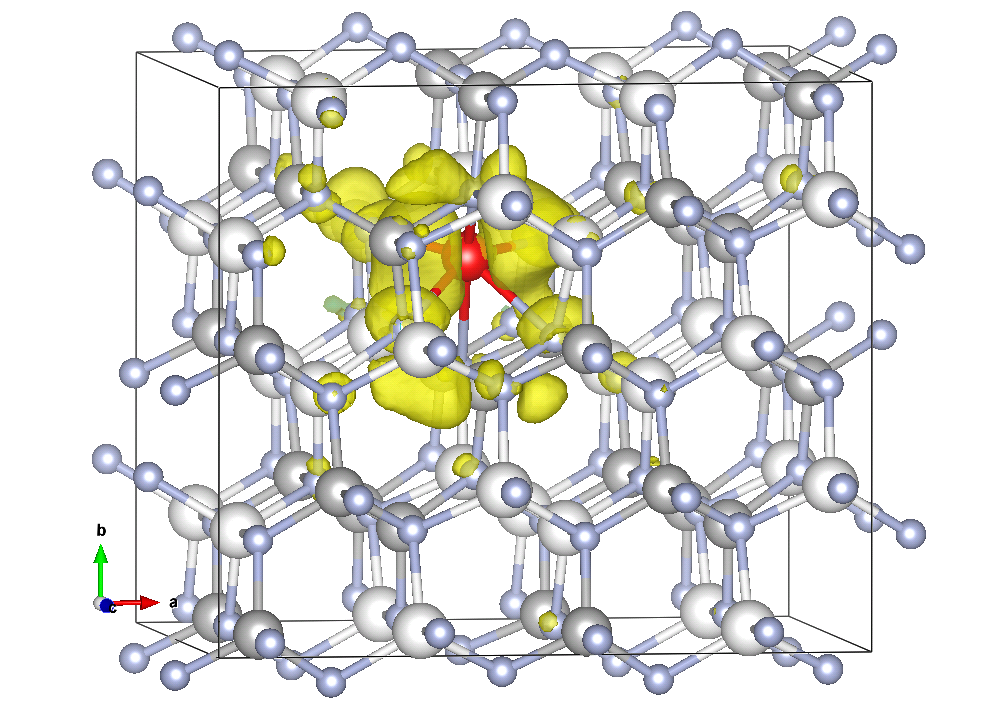}}
\caption{(Color online) Wave functions $|\psi|^2$ of a few 
selected states at ${\bf k}=0$  for (a) HOMO (highest occupied molecular orbital) corresponding to level in the gap, and resonances (b) HOMO+2 and (c) HOMO+3 in Ge$_{i}$ in the $q=+2$ charge state. Red sphere indicates Ge$_{i}$ atom. 
\label{wden_Ge_i_q_2}}
\end{figure}

Next, we consider the Ge$_i$ bands and PDOS in Fig. \ref{bnds_Ge_i} within LDA+U.  
We show it again for the $q=+2$ charge state. In contrast to the Zn$_i$,
we now see a sharp defect level about 1.5 eV above the VBM. For the
neutral charge state, again there would be electrons in the CB but now
the $q=+3$ and $q=+4$ charge states also could occur by depleting
the defect level in the gap. Fig.\ref{wden_Ge_i_q_2} illustrates
some of the relevant wave functions. These orbital shapes are almost the same in
LDA or LDA+U. They are here shown for LDA.
The HOMO of the ${q=+2}$ charge
state corresponds to the defect level in the gap and is clearly
localized around the Ge interstitial. However, the HOMO+2 and HOMO+3
above the conduction band indicate there are also resonances in the
conduction band.  The deep level in the gap has mostly Ge-$s$ character as
can be seen from its more or less spherical looking charge density
contribution on the Ge site. In contrast, the resonances in the conduction
band have Ge-$p$ character. Both defect states are seen to also have
some admixture of neighboring N-$p$ states.
Comparing to LDA, (see Supplementary Information) we find that the defect level in the gap
  moved  up by the LDA+U correction from 0.53 eV to 1.40 eV. This shift of $\sim$0.9 eV should
  be compared with the gap shift of 1.2 eV. It shows that although
  this level lies deep below the CBM it is still strongly
  shifting along with the CBM, which is consistent with its Ge-$s$ character.
  This shift also affects the positions of the transition levels, shown in Fig. \ref{E_for}. For the latter, we note that an effective charge of  $q_\mathrm{eff}=+2.3$
  was used for the $+3$ charge state to best match the distance dependence of
  the electrostatic potential. The choice of $q_\mathrm{eff}$ affects the
  position of the transition state. We estimate that this may lead to
  an uncertainty of order a few 0.1 eV on these transition levels. 

 \begin{figure}[h]
\includegraphics[width=5cm]{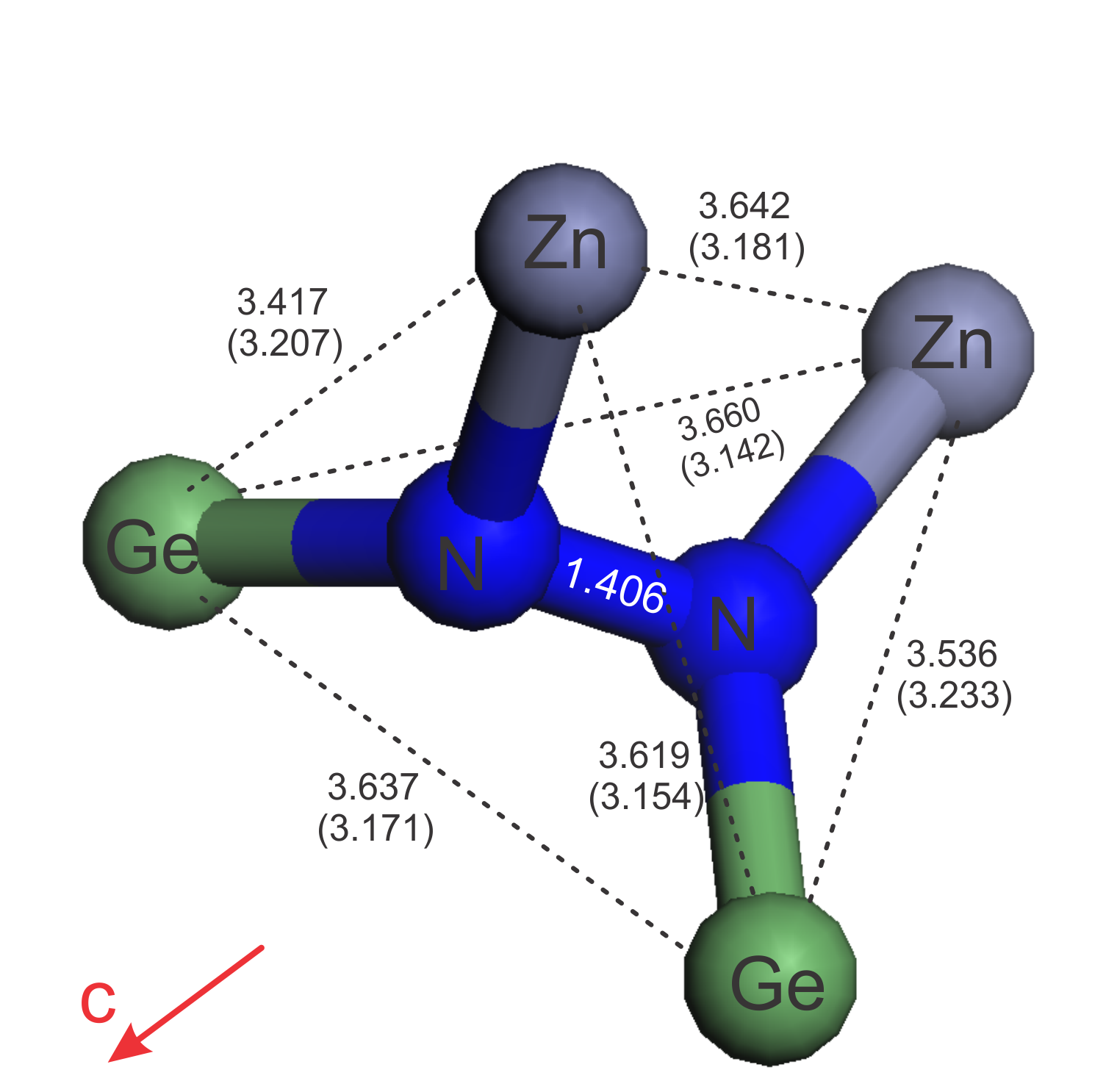}
\includegraphics[width=5cm]{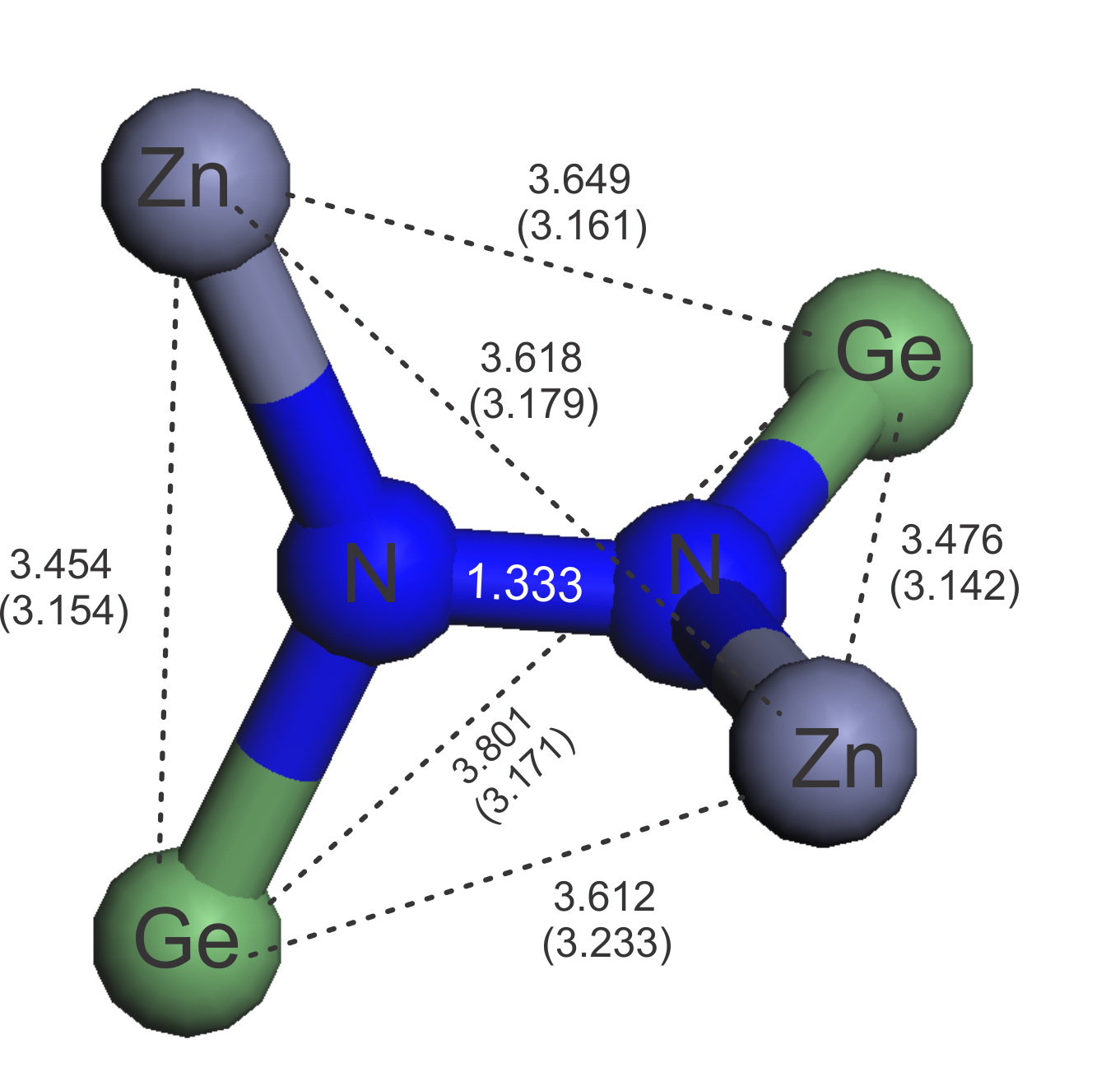}
\caption{(Color online) Geometry of two different types of N$_{i}$ complexes, \textit{i}1 (top) and \textit{i}2 (bottom). The numbers are the length in \AA\ 
  corresponding to the neutral charge state of the defect.
  The numbers in parentheses are the lengths in perfect crystal.
\label{Geometry_N_i}}
\end{figure}

The interstitial N$_i$  turns out to have a more interesting relaxation
behavior than the Zn$_i$ or Ge$_i$ which show only small displacements
of the neighboring atoms. In fact, the N$_i$ forms a split-interstitial
configuration where a N$_2$ dumbbell replaces the $N$-site. In other
words, the N interstitial displaces one of the neighboring N lattice sites
by forming a bond with it.  This is shown in Fig. \ref{Geometry_N_i} which
shows the bond-lengths for the two slightly different N interstitial
configurations we found. The first one, $i1$ corresponds to
N$_2$ replacing the N$_\mathrm{Zn}$ which sits above a Zn, atom
in the $c$-direction. The second one, $i2$ corresponds to a N$_2$
sitting above a Ge. The latter has a slightly shorter N$_2$ bond length
but both are larger than in a free N$_2$ molecule, which has
a bond length of $\sim$1.10~\AA.

\begin{figure}
\subfigure[]{\includegraphics[width=8cm]{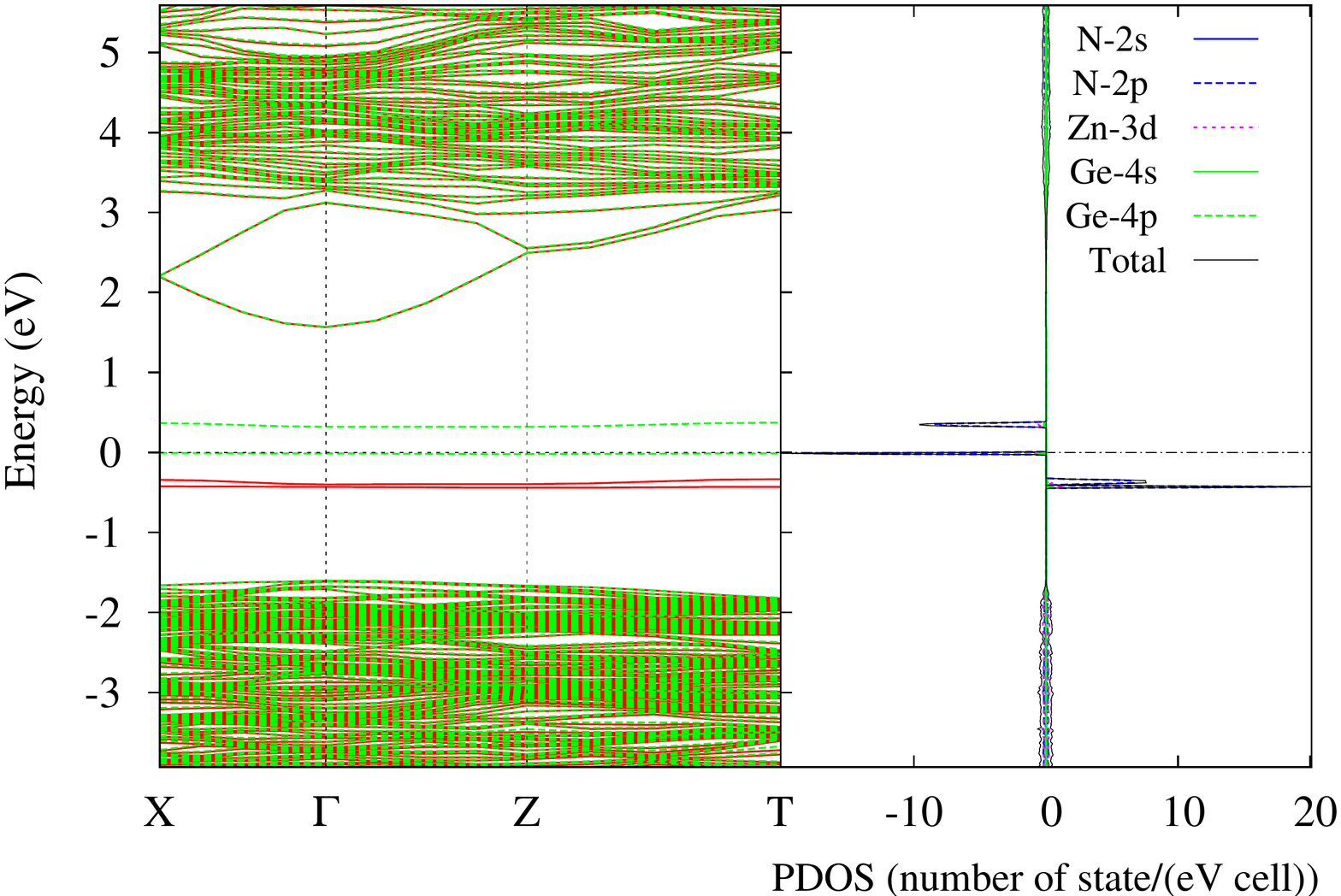}}\vspace{-1cm}
\subfigure[]{\includegraphics[width=8cm]{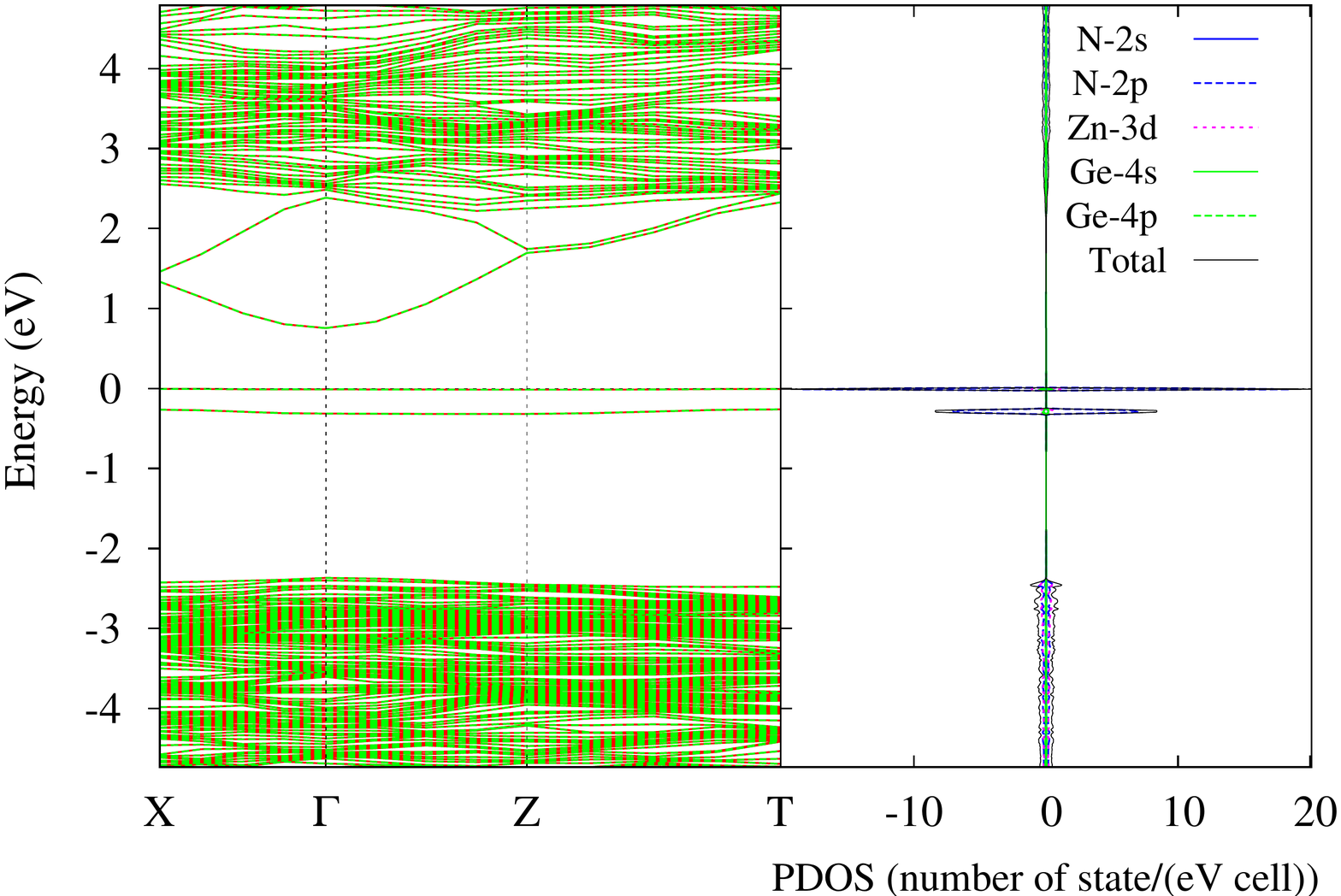}}\vspace{-1cm}
\subfigure[]{\includegraphics[width=8cm]{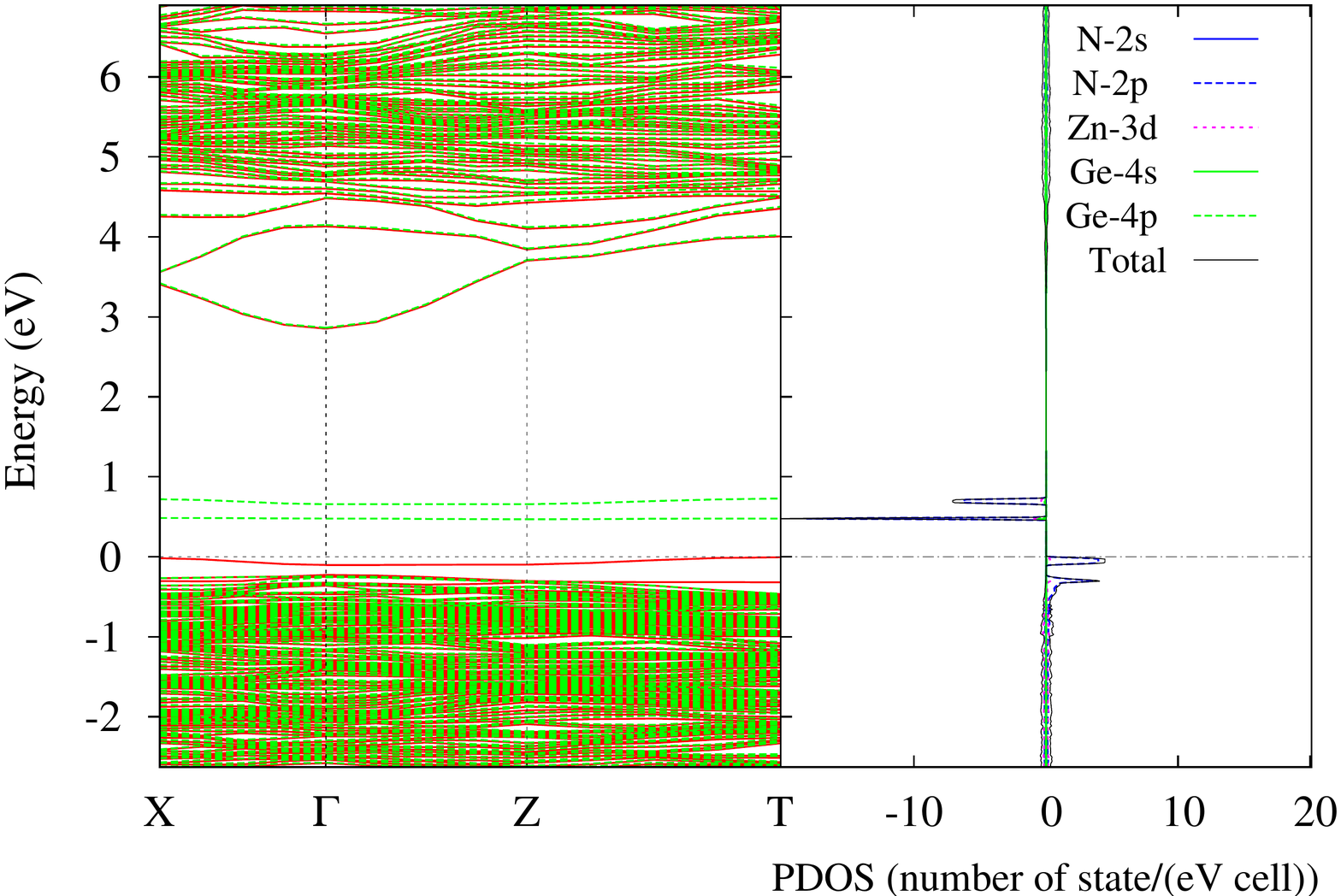}}
\caption{(Color on-line) LDA+U band structure and projected density of states on nearest atoms 
  to the ${\rm N}_{\rm i2}$ in (a) $q=0$, (b) $q=-1$, and (c)
  $q=+1$  states. Red and green lines indicate majority and minority spin bands. \label{bnds_N_i}
}
\end{figure}

\begin{figure}[h]
 \includegraphics[width=8cm]{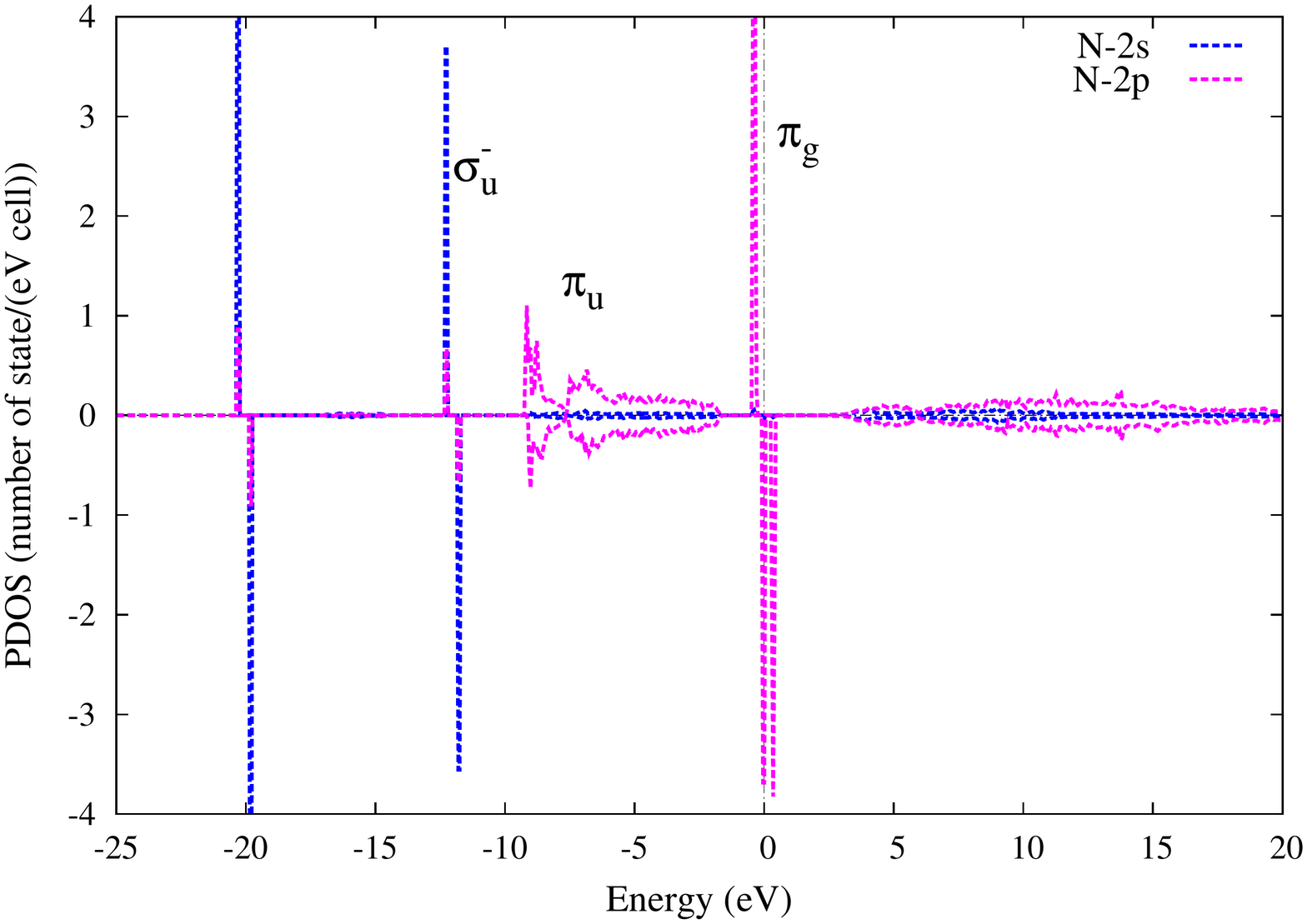}
 \caption{(Color online) 
Density of states of N$_2$ complex.\label{PDOS_N2}}
\end{figure}

Fig. \ref{bnds_N_i} shows the energy levels and PDOS for
the N$_{i2}$ configuration  as an example and obtained within LDA+U
and including spin-polarization. 
(Results within LDA are given in Supplementary Information.)
The results for the other one are similar.  We can see two closely spaced
defect levels in the gap for each spin.   Fig. \ref{PDOS_N2} shows the N$_2$ projected
PDOS on a wider energy scale, from which we can identify the molecular
orbitals of the N$_2$ molecule, as indicated by the labeling.  It shows
that the states in the gap correspond to the $\pi_g$ state, which in
the N$_2$ molecule would be the lowest unoccupied molecular orbital
(LUMO).  As expected, since this N$_2$ molecule now plays the role
of a N ion in the crystal, it accommodates three extra electrons,
and thus the neutral charge state of the defect, in a nominal sense
corresponds to the N$_2^{3-}$ state of the molecule. Of course, the
charge will be spread out somewhat to neighboring atoms.  We can also
see this from Fig. \ref{bnds_N_i}.
In fact, in the neutral charge state, the Fermi level
coincides with the lower minority spin level, thus the levels are occupied with 3 electrons. 
In the LDA and non-spin-polarized calculation, the level passes through the upper defect level
band.  On the other hand, in the $q=-1$ state all levels in the gap are filled and a $S=0$ state
results. Finally, in the $q=+1$, we have two electron in the lower levels. They can in principle have a
$S=0$ or $S=1$ state. We find the latter to have the lower energy, resulting in a high-spin state of the defect. 
When this state is further emptied, a $q=+2$ state with $S=1/2$ is also found to give a transition level in the gap. 

\begin{figure}[h]
\includegraphics[width=8cm]{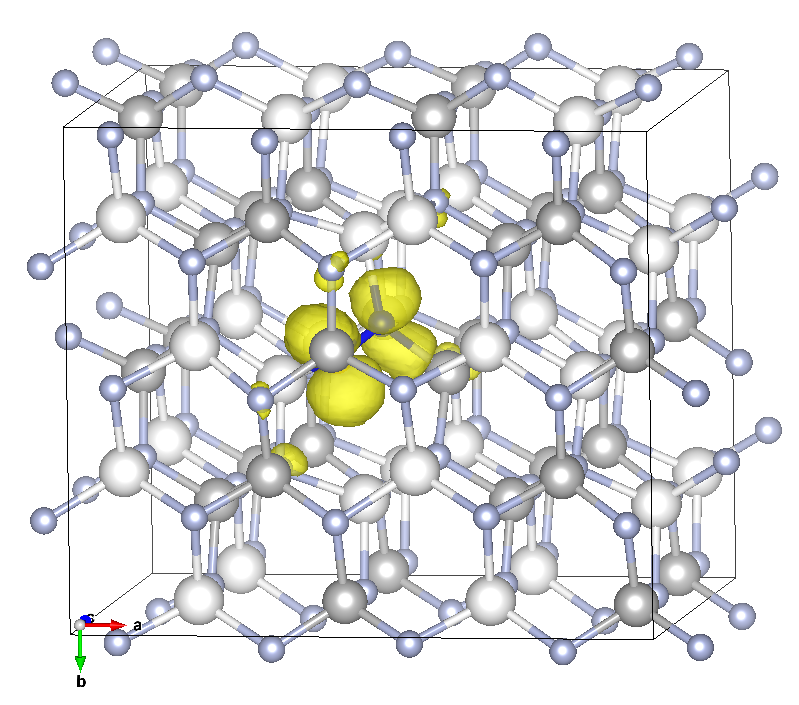}
\caption{(Color online) Wave function modulo squared for the HOMO
  or highest level in the gap 
  in N$_{i2}$ in the $q=0$ charge state. The white and gray spheres indicate Zn and Ge atoms, respectively. The light blue spheres indicate regular N atoms, whereas dark blue indicate N$_{2}$ complex.  The next level HOMO-1 looks very similar but is not shown.
\label{wden_N_i2_q_0}}
\end{figure}

The energies of formation
in Fig. \ref{E_for} show  the corresponding transition levels $2+/1+$, $1+/0$ and
$0/-1$.  The occurrence of both +1 and $-1$ charge states, shows that
this is an amphoteric trap level: it can both catch holes and electrons. 
The wave function plot in Fig. \ref{wden_N_i2_q_0} for the neutral
charge states corresponds to the upper level
and confirms the N$_2$ $\pi$-like molecular
state character  of this level. The second level looks almost identical
and is therefore not shown.
The slight splitting of this level
results from the interaction with the crystalline environment which
lowers the symmetry compared to the free molecule and hence,
splits the 2-fold orbital degeneracy of the $\pi$-state.
For the $N_{i1}$ interstitial, the results are similar but the defect levels
split slightly more. This difference corresponds to the slightly different
bond distances of the N atoms to the neighboring Zn and Ge.

Comparing with the LDA approach we find that the N$_i$ defect levels 
  stay at approximately the same energy relative to the VBM (they
  move up by only $\sim$0.2 eV due to the $U$ terms) while the
  CBM moves up significantly.   This is consistent with the deep and N-$2p$
  character of this
defect.  On the other hand, including the spin-polarization effects of these levels is important.

In Table \ref{tabefor} we give the energies of formation, corresponding
to the chemical potential condition C.  
Compared with the energies of formation of the vacancies and
antisites, which were studied in Ref. \onlinecite{Skachkov16},
we see that the interstitials have higher energy of formation than
the antisites.  They thus will play little role in the
determination of the Fermi level of the native material, which
is dominated by the low energy of formation antisite competition.
While these defects, having high energy of formation are thus
not expected to occur in large concentrations in equilibrium, they
are nonetheless important because they could be formed under
irradiation or ion bombardment.  The transition levels are
summarized in Table \ref{tabtrans}.


\begin{table}[h]
\centering
\caption{Formation energy (in eV) for various defects in different 
  charge states at $\epsilon_F=0$. The calculation
  corresponds to chemical potentials of point C in 
  Table \ref{tabchempot}. LDA+U corrections of the occupied one-electron defect levels
  are included. 
  \label{tabefor}}
\begin{ruledtabular}
\begin{tabular}{c c c } 
Defect & $q$ &   $E_{\rm for}(\epsilon_F$=0) \\
\hline
Zn$_{\rm i}$  &  2 &      4.51  \\ \hline
Ge$_{\rm i}$  &  2 &      6.16  \\
              &  3 &      5.07  \\ 
              &  4 &      4.69  \\ \hline
N$_{\rm i1}$   &  0 &     5.14  \\ 
               &  1 &     4.40  \\ 
               &  2 &     4.05  \\ 
               & -1 &     7.49  \\ \hline
N$_{\rm i2}$   &  0 &     5.33 \\ 
               &  1 &     4.61  \\ 
               &  2 &     4.28  \\ 
               & -1 &     7.52 \\  
\end{tabular}
\end{ruledtabular}
\end{table}

\begin{table}[h!]
\centering
\caption{Transition levels above the VBM in LDA+U (in eV). \label{tabtrans}
}
\begin{ruledtabular}
\begin{tabular}{c c c c } 
 Defect &\multicolumn{2}{c}{ Transition levels} \\
 & +2/+1 & +1/0 & 0/-1  \\ [0.5ex] 
 \hline 
 N$_{\rm i1}$ & 0.34 & 0.74 & 2.35    \\
 N$_{\rm i2}$ & 0.33 & 0.72 & 2.19    \\
\hline
 & +4/+3 & +3/+2 &   \\ [0.5ex] 
 \hline 
 Ge$_{\rm i}$ &  0.38 & 1.09      \\
\end{tabular}
\end{ruledtabular}
\end{table}

\section{Conclusions}
In this paper, we studied the native interstitial defects in
ZnGeN$_2$. We found that both Zn$_i$ and Ge$_i$ defects behave
as double shallow donors and prefer the octahedral
interstitial site because of their size.
However, Ge$_i$ also has a deep level close
to the VBM and can thus occur in a $+3$ and even $+4$  charge state.
The level in the gap is Ge-$s$ like. Both defects have resonances in the
CB which in the Zn$_i$ case corresponds to Zn-$s$ but in the Ge$_i$ case
corresponds to Ge-$p$.  The N$_i$ forms two slightly different
split interstitials depending on its location, with slightly
different effective N$_2$ bond lengths. It has a level in the gap
corresponding to the $\pi$ like LUMO of the N$_2$ molecule and
this level behaves amphoteric as a trap level, in other words, it
can be both in positive and negative charge states. In its single positive charge,
it prefers a high-spin $S=1$ state. 
The interstitials are all found to have higher energies of formation
than the antisites, studied earlier and will hence not affect
the position of the Fermi level because of their expected low
concentration in equilibrium. In particular, they are not expected
to contribute significantly to the background unintentional n-type doping
found in most ZnGeN$_2$ crystals.

\acknowledgements{This work was supported by the U.S. National 
Science Foundation under Grant No. DMREF-1533957. The calculations were performed at the 
Ohio Super Computer Center under Project No. PDS0145.}

\bibliography{zgn,zsn,lmto,gw,dft,defects,msn,ldau}

\end{document}